\documentclass[%
preprint,
 amsmath,amssymb,
 aps,
prb,
]{revtex4-2}

\usepackage{graphicx}
\usepackage{dcolumn}
\usepackage{bm}
\usepackage [english]{babel}
\usepackage [autostyle, english = american]{csquotes}
\usepackage{subfiles}
\usepackage{textgreek}

\begin{document}

This manuscript has been authored by UT-Battelle, LLC under Contract No. DE-AC05-00OR22725 with the U.S. Department of Energy. The United States Government retains and the publisher, by accepting the article for publication, acknowledges that the United States Government retains a non-exclusive, paid-up, irrevocable, world-wide license to publish or reproduce the published form of this manuscript, or allow others to do so, for United States Government purposes. The Department of Energy will provide public access to these results of federally sponsored research in accordance with the DOE Public Access Plan(http://energy.gov/downloads/doe-public-access-plan).

\title{Collective nature of phonon energies in solids\\ beyond harmonic oscillators}

\author{Jaeyun Moon}
\email{To whom correspondence should be addressed; E-mail: jaeyun.moon@cornell.edu}
 \affiliation{Sibley School of Mechanical and Aerospace Engineering\\ Cornell University, Ithaca, New York, USA, 14853}

\author{Leo Zella}
 \affiliation{Physical and Life Sciences Directorate,\\ Lawrence Livermore National Laboratory, Livermore, California, USA, 94550}
\author{Lucas Lindsay}
 \affiliation{Materials Science and Technology Division,\\ Oak Ridge National Laboratory, Oak Ridge, Tennessee, USA, 37831}

\date{\today}
\clearpage

\begin{abstract}
 Phonon quasi-particles have been monumental in microscopically understanding thermodynamics and transport properties in condensed matter for decades. Phonons have one-to-one correspondence with harmonic eigenstates and their energies are often described by simple independent harmonic oscillator models. Higher order terms in the potential energy lead to interactions among them, resulting in finite lifetimes and frequency shifts, even in perfect crystals. However, increasing evidence including constant volume heat capacity different from the expected Dulong-Petit law suggests the need for re-evaluation of phonons having harmonic energies. In this work, we explicitly examine inter-mode dependence of phonon energies of a prototypical crystal, silicon, through energy covariance calculations and demonstrate the concerted nature of phonon energies even at 300 K, questioning independent harmonic oscillator assumptions commonly used for phonon energy descriptions of thermodynamics and transport.

\end{abstract}


\maketitle
\clearpage

\textit{Introduction.}-- In crystalline solids, quanta of atomic vibrations, or phonons, have been used to microscopically describe thermodynamics and transport properties for over a century with much success \cite{debye_zur_1912, einstein_plancksche_1907}. Building upon this, engineering phonon properties such as frequencies and lifetimes have been a focus in design principles for controlling macroscopic thermal properties of solids from a bottom-up perspective \cite{qian_phonon-engineered_2021, hanus_thermal_2021}. 

Phonons are often described synonymously as normal modes of harmonic oscillators in various textbooks \cite{ashcroft_solid_1976, kittel_introduction_1976} and are obtained by diagonalizing dynamical matrices describing the equations of motion for the oscillators. These involve second order interatomic force constants (second derivatives of the potential energy) and oscillator masses, typically atoms. The resulting harmonic modes are non-interacting and orthonormal. Mode frequency as a function of wavevector gives phonon dispersions and mode group velocities. Summing over the thermally populated harmonic oscillator energies and zero-point energy gives the total vibrational energy, whose temperature derivative is the total heat capacity. By including anharmonicity in the potential energy surface, one can describe interactions between phonons, leading to finite intrinsic phonon lifetimes, even in perfect crystals \cite{maradudin_scattering_1962}. Because lifetimes can vary by orders of magnitude depending on temperature, frequency, and material, engineering phonon scattering has great promise, driving efforts to understand various interaction mechanisms including phonon-phonon scattering, phonon-defect scattering, and tunneling \cite{hanus_thermal_2021, simoncelli_unified_2019,allen_thermal_1989, deangelis_thermal_2018}. This framework has led to many successful heat capacity and thermal conductivity predictions with respect to measurements in various solids \cite{lindsay_survey_2018, mcgaughey_predicting_2014, mcgaughey_phonon_2019}. 

For more precise predictions, particularly in strongly anharmonic solids, some have considered how the anharmonic interactions give rise to phonon renormalization, leading to frequency shifts, further affecting the lifetimes and group velocities measured by the frequency broadening and slopes of the renormalized phonon dispersions, respectively \cite{hellman_temperature_2013, errea_anharmonic_2014, monacelli_stochastic_2021}. In these schemes, renormalized phonons are, however, often effectively considered to have independent harmonic energies where phonon energies are a function of their modes only: in the overall thermal conductivity in terms of phonons, $k = \sum_i c_i \alpha_i$ where $c_i = \hbar\omega_i \text{d}n/\text{d}T$ is the anharmonically renormalized mode heat capacity, $\alpha_i$ is the mode thermal diffusivity describing phonon interactions \cite{hellman_phonon_2014, pandit_effect_2021}, and $n$ is the Bose-Einstein statistics. In the classical limit, renormalized specific heat reaches the Dulong-Petit law. However, growing evidence from MD simulations and theoretical approaches demonstrates specific heat values departing from the Dulong-Petit law in various crystals and amorphous solids under constant volume \cite{allen_anharmonic_2015, andritsos_heat_2013, moon_normal_2024}, suggesting the need for re-evaluation of phonons having simple harmonic oscillator energies. 

In this work, we investigate the effect of anharmonicity in renormalized individual mode energies in a prototypical semiconductor, silicon, via molecular dynamics (MD) and lattice dynamics (LD) from 300 to 1500 K. By examining the self-variance of each mode energy and covariance with other mode energies, we find that phonons are  \textit{dependent} on other mode energies for all modes at all temperatures with increasing coupling as temperature increases. Consequently, our results demonstrate that mode energies are intrinsically collective in nature and challenges many assumptions regarding phonon thermodynamics and transport theories that assume independent phonon energies.

\textit{Formalism.} -- 
The Hamiltonian of a system of harmonic oscillators can be expressed in terms of normal mode coordinates, $Q_i(t)$, and their time derivatives, $\dot Q_i(t)$ as
\begin{equation}
    H(t) = \sum_i H_i(t) = \sum_{i}  \underbrace{\frac{1}{2}\omega^2_i Q_i(t)^2}_\text{PE\textsubscript{i}} +  \sum_{i} \underbrace{\frac{1}{2}\dot Q_i(t)^2}_\text{KE\textsubscript{i}} 
    \label{eq: hamiltonian}
\end{equation}
where $\omega_i$ is the renormalized mode radial frequency. The first term and the second term in the above Hamiltonian expressions are the potential energy (PE) and kinetic energy (KE), respectively. Normal mode coordinates are related to atomic displacements, $\boldsymbol{u}_{n}(t)$, as
\begin{equation}
Q_i(t) = \sum_{n}m_n^{\frac{1}{2}}\boldsymbol{e}_{i,n} \cdot \boldsymbol{u}_{n}(t)
\end{equation}
where $m_n$ is the atomic mass and $\boldsymbol{e}_{i,n}$ is the eigenvector of the second order dynamical matrix. We consider entire simulation domains as a large unit cell and therefore only consider unfolded $\Gamma$ point modes for simplicity and without loss of generalization \cite{dove_introduction_1993, moon_heat_2024}. Despite the harmonic formalism in Eq. \ref{eq: hamiltonian}, displacements within MD at finite temperature, which account for all degrees of anharmonicities, are projected to the normal mode coordinates such that anharmonicity is intrinsically included. 

The temporal decay of the autocorrelation of $H_i$ is then related to the relaxation time of each mode \cite{ladd_lattice_1986,mcgaughey_phonon_2006}. Renormalized or anharmonic mode frequency ($f_{anhar}$) and lifetime ($\tau$) can be extracted by fitting either normalized autocorrelation of PE\textsubscript{i} or KE\textsubscript{i} by $\text{cos}^2(2\pi f_{anhar}t)e^{-t/\tau}$. We find that fitting to both autocorrelations leads to the same frequencies and lifetimes. For $\omega_i$ in Eq. \ref{eq: hamiltonian}, we have used anharmonic radial frequencies ($2\pi f_{anhar}$) to re-compute mode potential energy. This scheme of phonon calculations have been used for many crystalline and amorphous solids in prior works for accurate thermal conductivity predictions against Green-Kubo (GK) calculations \cite{mcgaughey_phonon_2006, parrish_origins_2014, zhou_quantitatively_2015, turney_predicting_2009, hashemi_effects_2020, feng_quantum_2016, ye_spectral_2015}. Multi-phonon interactions are intrinsically accounted for in this formalism, thus normal mode lifetimes are in agreement with perturbation theory predictions including three- and four-phonon processes for various crystalline solids, including silicon. Furthermore, these normal mode lifetimes have led to consistent thermal conductivity calculations with those from GK up to high temperatures \cite{feng_quantum_2016}.

Here, molecular dynamics simulations of silicon (1728 atoms) using the Stillinger-Weber potential \cite{stillinger_computer_1985} were carried out with large-scale atomic/molecular massively parallel simulator (LAMMPS) \cite{plimpton_fast_1995}. At each temperature, the system was equilibrated for 500 ps followed by 1 ns of data recording run with an atomic velocity and position sampling interval of 5 fs at each temperature from 300 K to 1500 K at constant volume (canonical ensemble). Timesteps of 0.5 fs were used. System size, sampling intervals and data recording times used here are similar to prior works to obtain phonon lifetimes and anharmonic frequencies \cite{zhou_quantitatively_2015, larkin_predicting_2013}. A lattice parameter of 5.431 \AA \ (0 K equilibrium lattice parameter) was used. For lattice dynamics calculations to obtain harmonic frequencies and eigenvectors from 0 K dynamical matrix, the general utility lattice program (GULP) was used \cite{gale_gulp:_1997}. The eigenvectors were then used to map the time dependent atomic trajectories and velocities to the normal mode coordinates and their time derivatives. 

\begin{figure}
	\centering
	\includegraphics[width=1\linewidth]{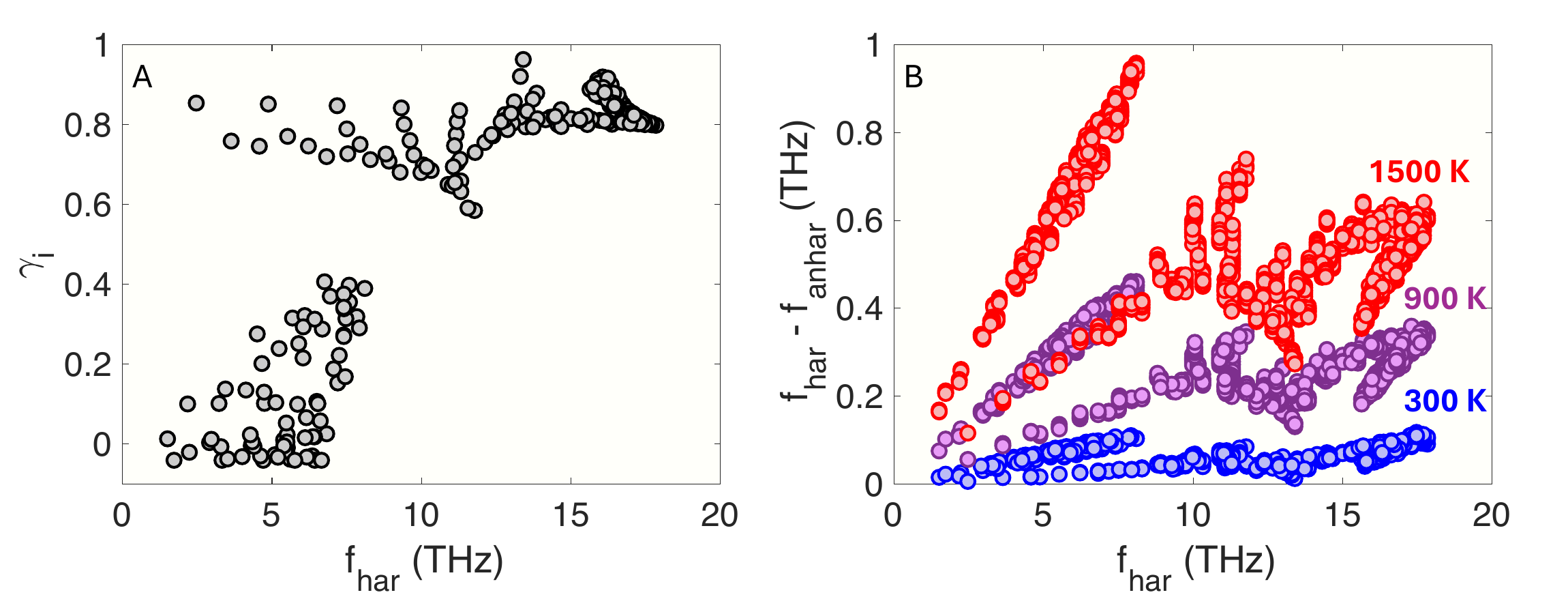}
	\caption{(A) Mode Gr\"uneisen parameters obtained by increasing the volume by 0.05 \%. Values and trends are similar to Ref. \cite{fabian_thermal_1997} using the same potential. (B) Temperature dependent individual phonon frequency shifts from 300 K to 1500 K }
	\label{fig:dos}
\end{figure}

\textit{Results and Discussion.} -- Mode Gr\"uneisen parameters given by $\gamma_i = - \text{d}ln(\omega_i)/\text{d}lnV$ are often used to describe mode level anharmonicity. Following the fitting procedures described above, anharmonic frequency shifts (relative to harmonic frequencies from LD calculations) are shown in Fig. \ref{fig:dos} along with mode Gr\"uneisen parameters where a volumetric change of 0.05 \% was used to obtain the derivatives in frequencies. Anharmonic frequency shifts from temperature dependent constant volume conditions are vastly different from Gr\"uneisen parameters. Higher frequency shifts are generally observed for higher temperatures.  Perhaps, the combination of these intrinsic (temperature) and volumetric effects can be used for a more universal mode anharmonic description \cite{allen_anharmonic_2015} and merit further studies into this matter.  

Frequency shifts shown in Fig. \ref{fig:dos} should be reflected in spectral velocity autucorrelation functions, which can be used to describe various types of atomic dynamics including atomic vibrations and diffusions \cite{moon_heat_2024, moon_sub-amorphous_2016}. In solids with no atomic diffusion, spectral velocity autocorrelations are often discussed synonymously with phonon density of states. However, these by themselves do not offer mode-level descriptions of the phonon behaviors. We observe decent agreement when comparing calculations of the phonon density of states from 300 to 1500 K from velocity autocorrelation functions and separately from our normal mode analysis using anharmonic frequencies (see Fig. S1), thus confirming our mode-level phonon renormalization is accurate.  

\begin{figure}[h!]
	\centering
	\includegraphics[width=0.7\linewidth]{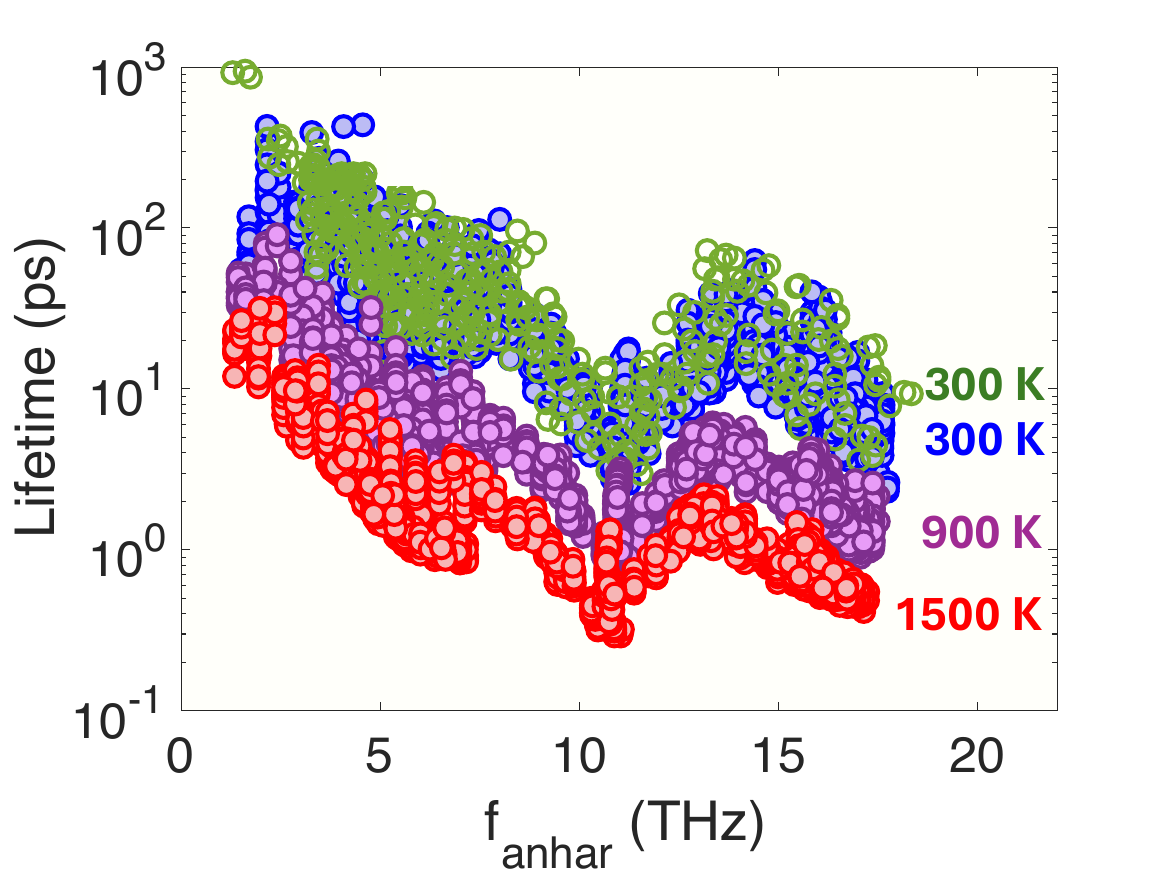}
	\caption{Temperature dependent phonon lifetimes at 300 (blue solid circles), 900 (purple solid circles), and 1500 K (red solid circles). Previously reported literature phonon lifetimes utilizing the same Stillinger-Weber potential are shown for reference (green open circles) \cite{larkin_predicting_2013}. }
	\label{fig:lifetime}
\end{figure}

Phonon lifetimes extracted from the fitting procedure discussed above are shown in Fig. \ref{fig:lifetime} from 300 to 1500 K. Our 300 K phonon lifetime values are in agreement with prior works using the same potential \cite{larkin_predicting_2013, zhou_quantitatively_2015}. As expected, a noticeable decrease in phonon lifetimes from 300 to 1500 K is demonstrated. For modes with the lowest lifetimes around 10 THz, lifetimes are still a few multiples of the period of the vibrations, and thus are reasonably defined quasi-particles. However, we anticipate that two-channel transport including phonon tunneling may be necessary at high temperatures \cite{simoncelli_unified_2019} for thermal transport calculations, especially for optical phonons where phonon dispersions are nearly overlapping above $\sim$ 10 THz \cite{babaei_machine-learning-based_2019}.

 With renormalized phonon frequencies and their respective mode energies, we next investigate how anharmonicities affect coupling between phonon energies. Variance of a linear combination of variables, like the anharmonic energies here, is given by 
 \begin{equation}
     \text{var}\Big(\sum_i X_i\Big) = \sum_i \text{var}(X_i) + \sum_{i\neq j} \text{covar}(X_i, X_j).
 \end{equation}
where $X_i = PE_i \ \text{or} \ KE_i$. For independent variables, the covariance contributions are zero and for dependent variables, covariance is non-negligible and can be positive or negative. Physically, a positive covariance between variables reveals that these variables tend to vary together in the same way. A negative covariance depicts an inverse relationship between the variables. Phonon energies are typically assumed to follow the harmonic oscillator model, $E_i = \hbar \omega_i(n+1/2)$, that is, an individual mode energy is a function of its mode only and is independent of the others \cite{larkin_thermal_2014, giri_atomic_2022, luo_vibrational_2020, moon_propagating_2018}. Therefore, we expect that covariance between individual mode energies is zero with some statistical fluctuations.

\begin{figure}
	\centering
	\includegraphics[width=1\linewidth]{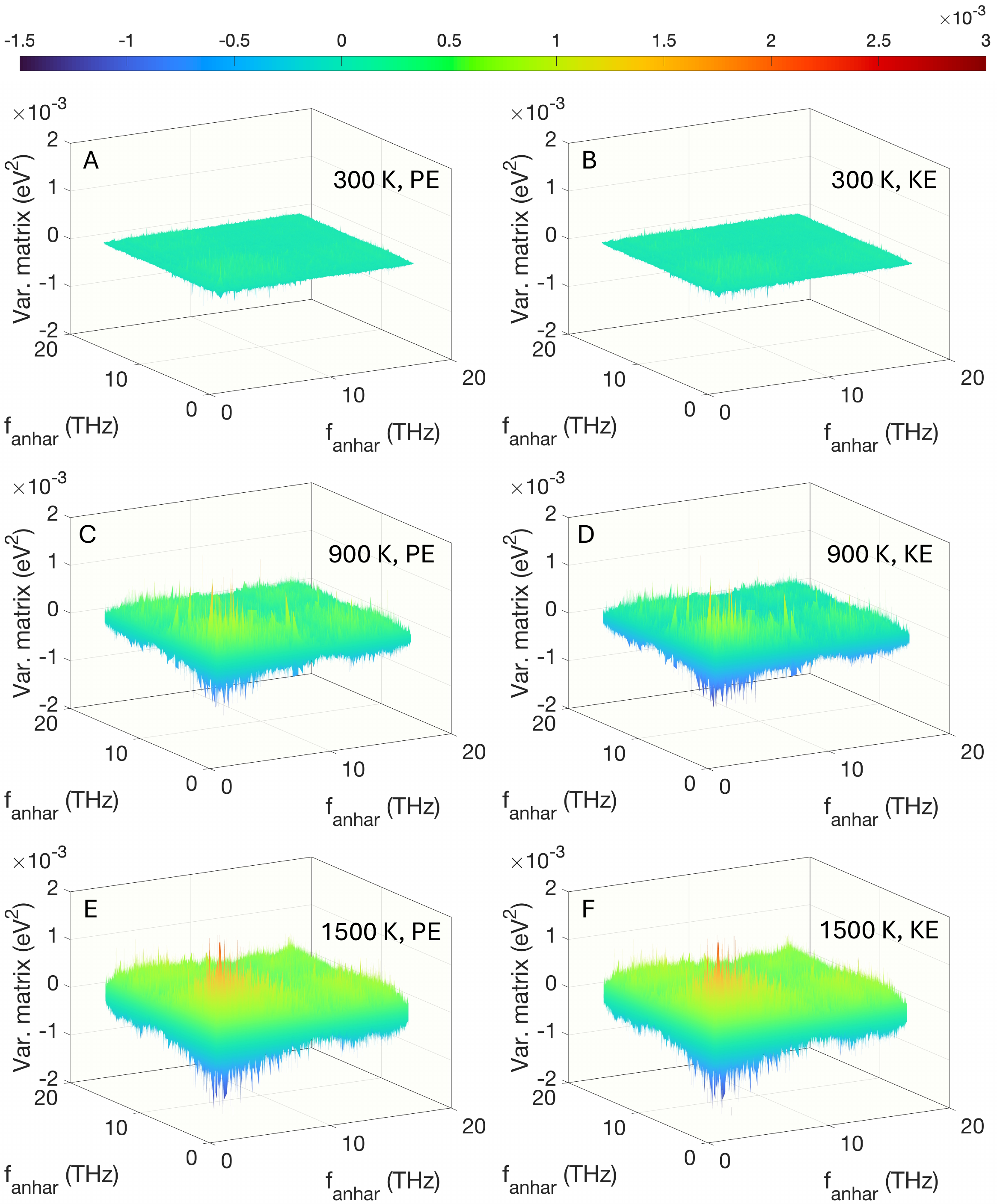}
	\caption{Energy covariance matrix for individual phonon modes at 300 (A, B), 900 (C, D), and 1500 K (E, F). Diagonal self-variance matrix together with covariance is shown in Fig. S2. Sub-figures on the left column and right column represent potential energy (PE) and kinetic energy (KE) variance matrix.}
	\label{fig:matrix}
\end{figure}

The covariance matrix of individual phonon mode energies without the self variance (i.e., var(X\textsubscript{i})) divided into potential and kinetic parts are shown in Fig. \ref{fig:matrix}. Total variance matrix including diagonal self-variance are plotted in Fig. S2. From the color map alone, potential and kinetic energy variance matrices appear nearly identical for each temperature. We observe non-negligible off-diagonal covariance fluctuations in both mode potential and kinetic energies even at 300 K (see Fig. S2). These fluctuations may arise from a combination of random thermal fluctuations and concerted interactions of phonons.  For more quantitative analysis, we sum over all mode contributions in variance for each mode and show diagonal and off-diagonal elements in Fig. S3 and S4 for 300 and 1500 K, respectively. 

Classically, due to atomic kinetic energy being $\frac{1}{2}m_n|\boldsymbol{v}_n|^2$ where $\boldsymbol{v}_n$ is the velocity of atom n, kinetic energy will always be $\frac{1}{2}k_BT$ per degree of freedom in the thermodynamic limit whereas potential energy in the thermodynamic limit changes depending on the anharmonicity. Therefore, in our finite simulations, both size and time, where we intrisically have thermal noise, comparing how the potential energy changes relative to the kinetic energy provides information about anharmonicity. For 300 K results, we observe nearly overlapping diagonal and off-diagonal elements for potential and kinetic energy variances. Off-diagonal covariances at 300 K appear to fluctuate near zero for both potential and kinetic energies without a clear pattern; however, the mean of the differences between the potential and kinetic energy covariances is non-zero. We anticipate that in the limit of infinite simulation time and number of atoms, the fluctuations will disappear while the mean remains. Therefore, we interpret the mean of the covariances as a measure of inter-mode dependence of phonon energies. Differences in diagonal elements between potential and kinetic energy variances are more noticeable at higher frequencies above $\sim$ 8 THz as shown in Fig. S3B.

For the 1500 K calculations, we observe clearer distinctions between the potential and kinetic energy diagonal variances (see Fig. S4). For harmonic oscillators, both potential and kinetic degrees of freedom depend linearly with temperature equally (equipartition theorem); therefore, equivalent variances between potential and kinetic mode energies from thermal fluctuations are expected. Indeed, nearly equal variances are observed for the lowest temperature calculations (300 K). However, we observe decoupling of the potential and kinetic energy variances, suggesting violation of the equipartition theorem for phonons here. We also observe that while the off-diagonal covariance fluctuates around zero for the kinetic energies, the mean of the off-diagonal covariance for potential energies have a finite offset value, signifying that phonon interactions through the potential energy surface lead to collective nature of phonon energies. We have additionally done time duration dependent self-variance and covariance energies and we observe that the means are nearly constant irrespective of time duration included in variance calculations while the spread becomes small as a function of increase in time duration as discussed previously (see Fig. S5). 

Genenrally, larger amplitudes in self variances and covariances are demonstrated for low frequency modes below $\sim$ 10 THz. We reason that because low frequency modes have larger lifetimes, their anharmonic interaction sampling is lower than higher frequency modes given the same time duration. However, these larger fluctuations at low frequencies do not affect our discussions.

\begin{figure}
	\centering
	\includegraphics[width=0.48\linewidth]{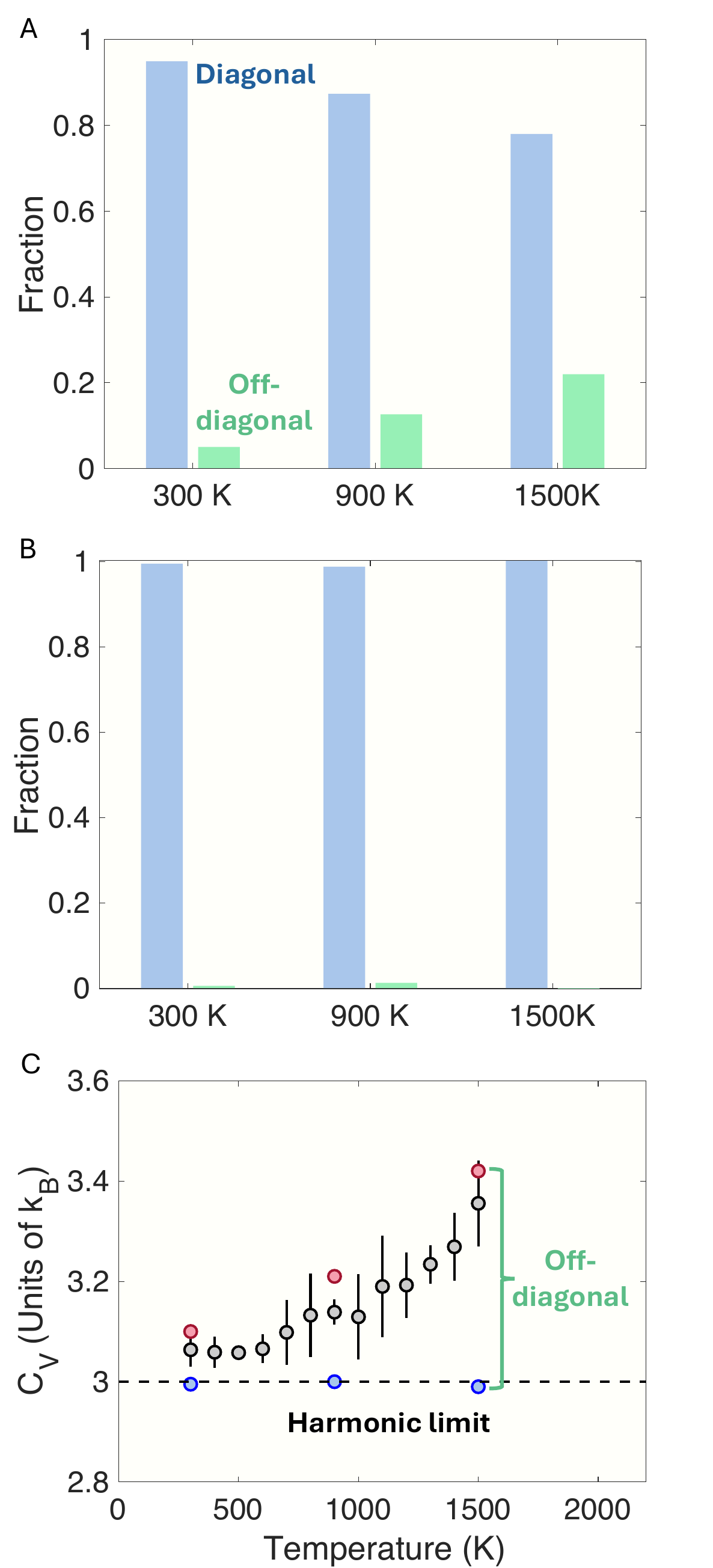}
	\caption{Respective contributions to the total variance in (A) potential and (B) kinetic energy. For the potential energy variance, co-variance elements (off-diagonal) representing inter-mode dependence increase in contributions with increase in temperature. For the kinetic energy variance, we observe negligible contributions from off-diagonal elements. (C) Heat Capacity comparisons from molecular dynamics simulations (black circles), kinetic energy total variances and self-variance of potential energy (blue circles), and total variances of potential and kinetic energies combined (red circles). Black dashed line is a guide to eye representing the harmonic oscillator limit. }
	\label{fig:fraction}
\end{figure}

Although individual diagonal elements have larger amplitudes compared to the off-diagonal elements, there are $3N(3N-2)$ more numbers of the off-diagonal elements than the diagonal elements where $N$ is the number of atoms. Therefore, net contributions to the total variance from off-diagonal elements could be non-negligible. Total contributions from self variance and covariances for each temperature are shown in Fig. \ref{fig:fraction}. We see that covariances in mode potential energies become increasingly significant with temperature while covariances remain negligible in mode kinetic energies. Our results explicitly demonstrate that phonon mode energies become inter-mode dependent and collective irrespective of frequency as evidenced by the lack of spectral dependence in off-diagonal covariance elements (see Fig. S3 and S4). Further, as the estimated effective zero-point energy temperature \cite{zella_ripples_2024} is $\sim$ 280 K for silicon, zero-point motion may be enough to cause these collective phonon energies in real silicon samples. Previously, zero-point energies have been shown to induce frequency shifts \cite{allen_anharmonic_2015}.

Our findings have direct consequences in existing thermodynamics and thermal transport theories. For instance, statistical mechanics in the canonical ensemble (as is in our simulations) dictates that total constant volume specific heat is linearly dependent on total energy variances \cite{mcquarrie_statistical_1976} as 
\begin{equation}
    C_V=\frac{\text{var}(E)}{k_BT^2}.
\end{equation}
Heat capacity from variance of atomic energies from molecular dynamics and variance from our phonon energy calculations are shown in Fig. \ref{fig:fraction}C. We observe decent agreement between these separate calculations. Inclusion of the mode kinetic energy variances and self-variance in the mode potential energies is approximately the harmonic Dulong-Petit limit of $3Nk_B$. However, these two contributions can also generally be anharmonic. This may be the case for other solids. Inclusion of the off-diagonal covariance of the mode potential energy describing intermode dependence leads to the observed temperature dependent anharmonicity in the specific heat, explicitly confirming the importance of collective interactions of phonon energies. A slight systematic difference in heat capacity from our phonon calculations and atomic energies may be due to higher order terms in the potential energy not included in Eq. \ref{eq: hamiltonian}. We anticipate that for some other solids such as Lennard Jones argon, the offset in the mean of the covariance of mode potential energy could be negative rather than positive shown here for silicon, We have shown previously that under constant volume with equilibrium density at 0 K, temperature dependent specific heat of crystalline argon is lower than the harmonic limit \cite{moon_normal_2024}.

In phonon transport calculations, it is often postulated that $k = \sum_i C_{V,i} \alpha_i$ where total specific heat is intrinsically assumed to be $C_V = \sum_i C_{V,i} = \sum_i \hbar \omega_i \text{d}n/\text{d}T$ with $n$ being the Bose-Einstein statistics. In this work, our mode energy self-variance and covariance analyses show the importance of off-diagonal inter-mode contributions and demonstrate that total specific heat is not the sum of individual mode contributions ($C_V \neq \sum_i \frac{\text{var}(E_i)}{k_BT^2}$) but rather including both individual and collective mode contributions is necessary, even when the lowest order is considered for the potential energy as in Eq. \ref{eq: hamiltonian}. We anticipate that higher order considerations in Eq. \ref{eq: hamiltonian} and the specific heat may be important in other solids, especially highly anharmonic solids such as perovskites. These effects will be considered in future works for a more complete treatment of the specific heat. We expect that our findings regarding the collective nature of phonon energies are generally applicable to various crystals and amorphous solids in both quantum mechanical and classical limits and also under other conditions such as constant temperature and pressure (NPT) where variance in enthalpy which is linearly dependent on energies is used to obtain constant pressure heat capacities.

\begin{figure}[h!]
	\centering
	\includegraphics[width=0.7\linewidth]{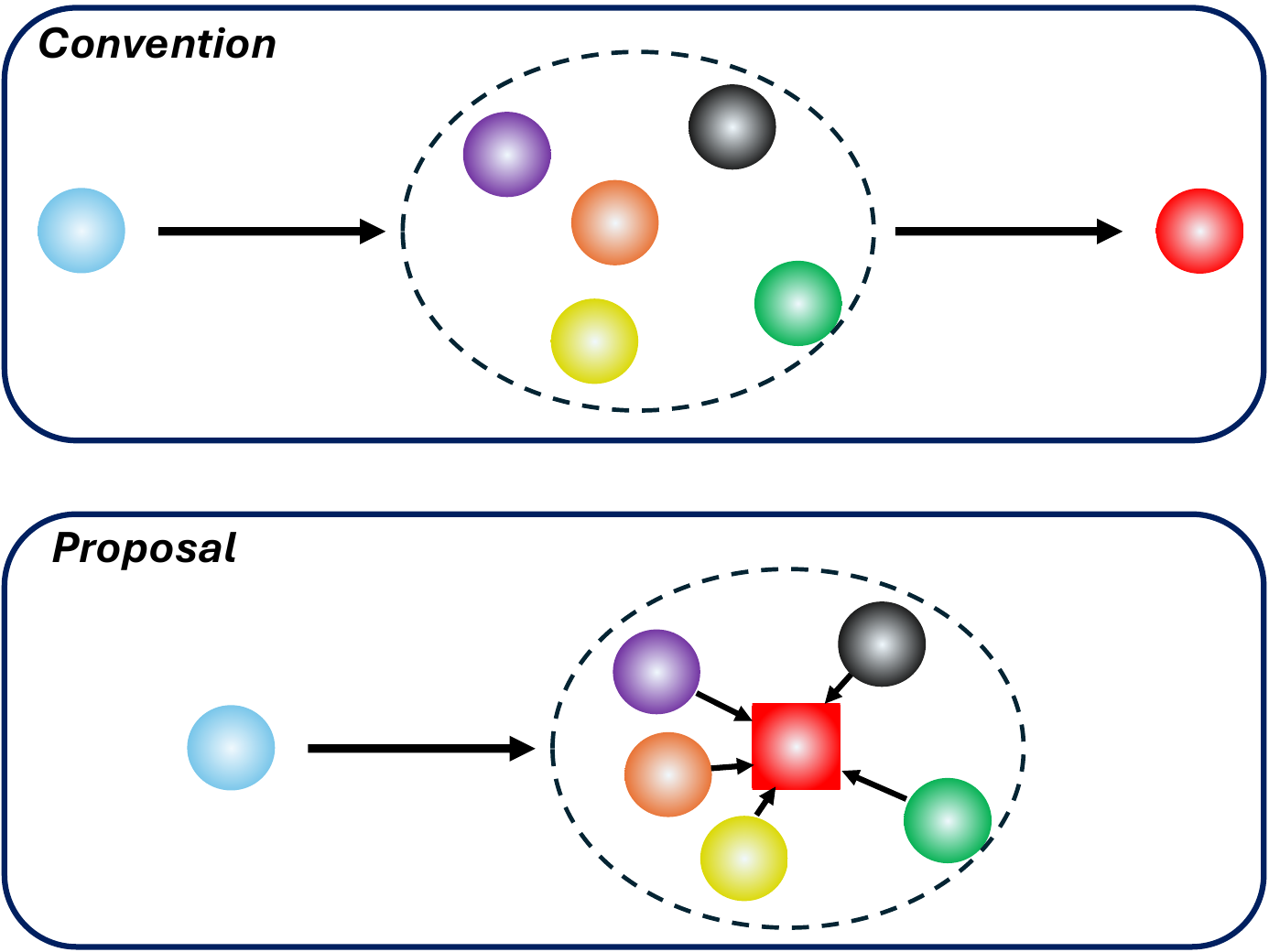}
	\caption{(A) Convention that phonon renormalization leads to frequency shifts in the harmonic oscillator model describing phonon energies as highlighted by color change where each sphere is a harmonic oscillator. (B) Our work proposes that phonon renormalization leads to the frequency shifts and phonon energies become collective beyond a simple harmonic oscillator (square).}
	\label{fig:schematic}
\end{figure}

\textit{Concluding remarks.} -- Phonon quasi-particle descriptions of atomic motion in solids have had immense success in describing thermodynamics and transport properties for past few decades where phonons are typically considered to have independent harmonic oscillator energies with frequencies shifted by anharmonic interactions as shown in Fig. \ref{fig:schematic}A. In this work, we explicitly demonstrate collective nature of phonon energies in a widely-studied semiconductor, silicon, even at room temperature as described in Fig. \ref{fig:schematic}B. Quantum mechanical fluctuations from zero-point motion may be enough to induce this behavior in real silicon samples. An example of consequences of our findings is described in anharmonic constant volume specific heat where inclusion of covariance of different phonon energies is necessary. Our results call into question single phonon descriptions of various thermodynamic and transport properties in solids.

\textit{Acknowledgement.} -- J.M. thanks Evan Willmarth at Yale for initial research discussions during his graduate internship at ORNL in 2023. We thank Philip Allen at Stony Brook University, Simon Th\'ebaud at INSA, and Xun Li at University of Texas, Austin for their comments and critical review of our manuscript. This work used the Advanced Cyberinfrastructure Coordination Ecosystem: Services \& Support (ACCESS) program which is supported by National Science Foundation grants \#2138259, \#2138286, \#2138307, \#2137603, and \#2138296 (Allocation PHY230148). L.L. acknowledges support by the U.S. Department of Energy, Office of Science, Office of Basic Energy Sciences, Material Sciences and Engineering Division for critical evaluation, discussions, and manuscript development. 

\section{Data Availability}
All data are included in the paper.

\clearpage

\clearpage

\title{Supplementary Information for\\
Collective nature of phonon energies in solids\\ beyond harmonic oscillators}
\author{Jaeyun Moon}
\email{To whom correspondence should be addressed; E-mail: jaeyun.moon@cornell.edu}
 \affiliation{Sibley School of Mechanical and Aerospace Engineering\\ Cornell University, Ithaca, New York, USA, 14853}

\author{Leo Zella}
 \affiliation{Physical and Life Sciences Directorate,\\ Lawrence Livermore National Laboratory, Livermore, California, USA, 94550}
\author{Lucas Lindsay}
 \affiliation{Materials Science and Technology Division,\\ Oak Ridge National Laboratory, Oak Ridge, Tennessee, USA, 37831}
 
\maketitle
\clearpage

\renewcommand{\figurename}{Fig.}
\setcounter{section}{0}
\renewcommand{\thesection}{\Roman{section}} 
\renewcommand{\thefigure}{S\arabic{figure}}
\setcounter{figure}{0}

\begin{figure}
	\centering
	\includegraphics[width=0.55\linewidth]{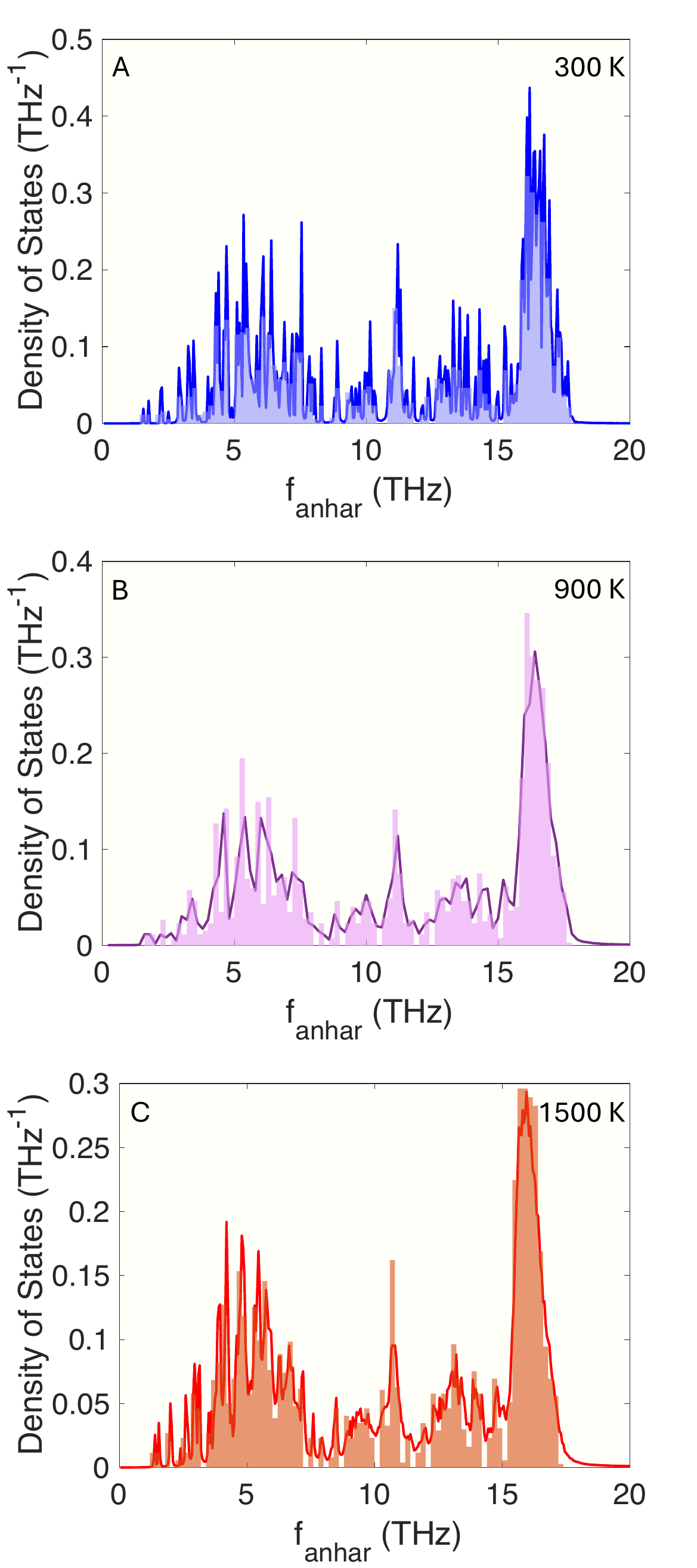}
	\caption{(A-C) Temperature dependent phonon density of states. Solid curves and shades represent velocity autocorrelation functions in frequency and population distribution of individual modes, respectively.}
	\label{fig:S1}
\end{figure}

\begin{figure}[h!]
	\centering
	\includegraphics[width=1\linewidth]{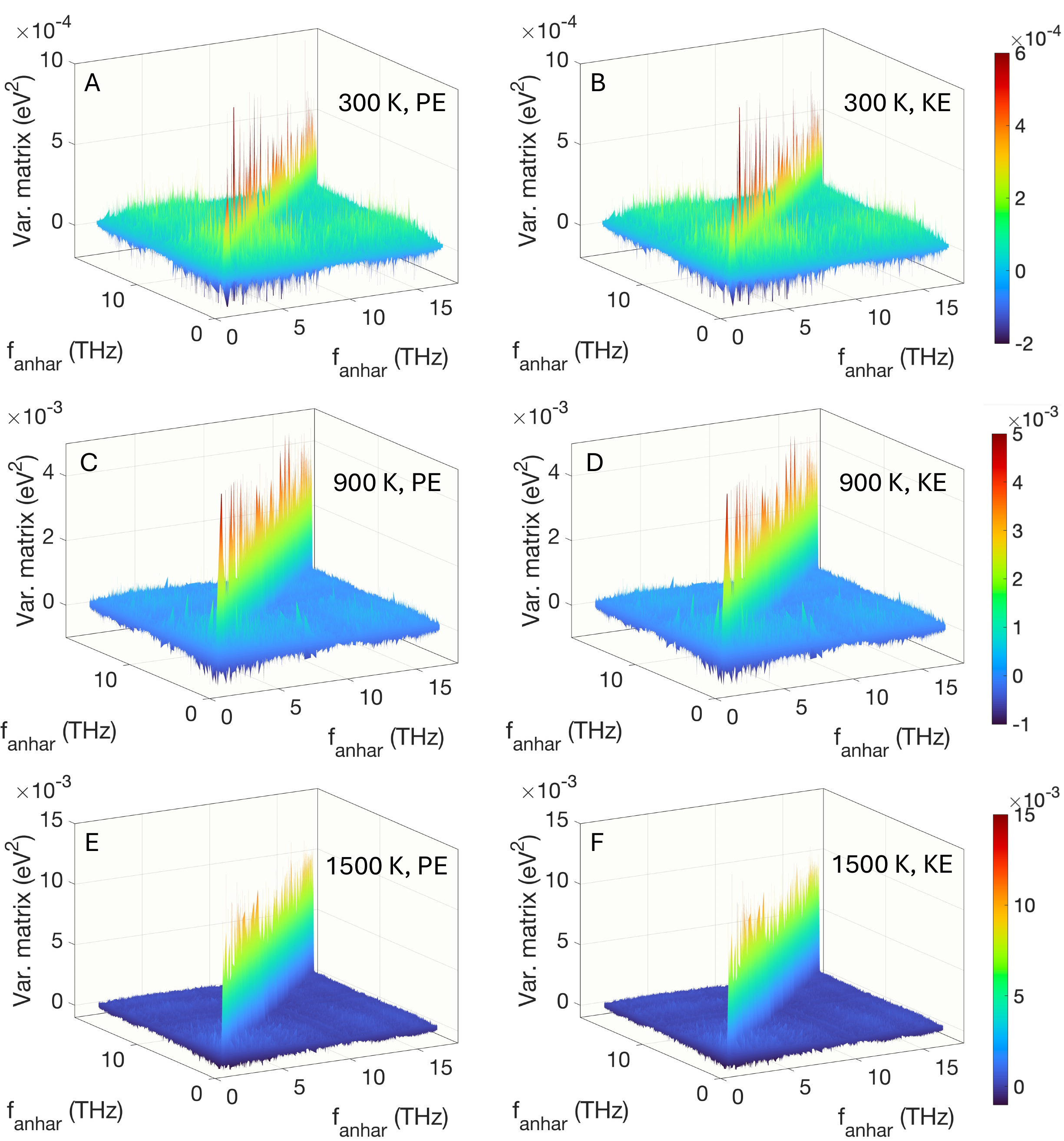}
	\caption{Energy variance matrix for individual phonon modes at 300 (A, B), 900 (C, D), and 1500 K (E, F). Sub-figures on the left column and right column represent potential energy (PE) and kinetic energy (KE) variance matrix.}
	\label{fig:S2}
\end{figure}

\begin{figure}[h!]
	\centering
	\includegraphics[width=1\linewidth]{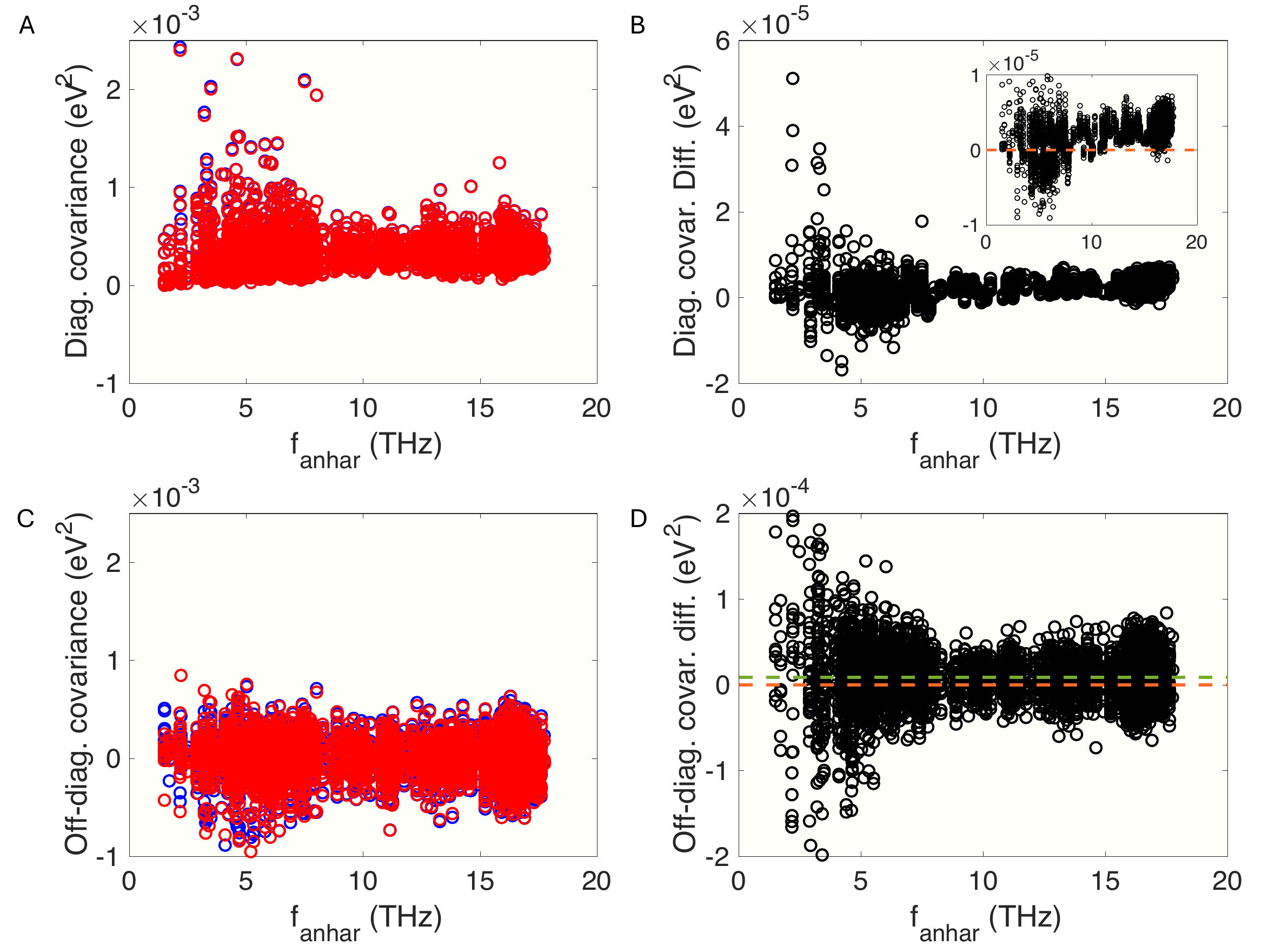}
	\caption{(A,C) Spectral self-variance and co-variance for mode potential (blue circles) and kinetic (red circles) energies, and (B, D) their differences (black circles) for 300 K. Inset in (B) is a zoomed-in view. For both (B, D), orange dashed lines represent zero and green dashed line is the actual mean.}
	\label{fig:S2}
\end{figure}

\clearpage

\begin{figure}[h!]
	\centering
	\includegraphics[width=1 \linewidth]{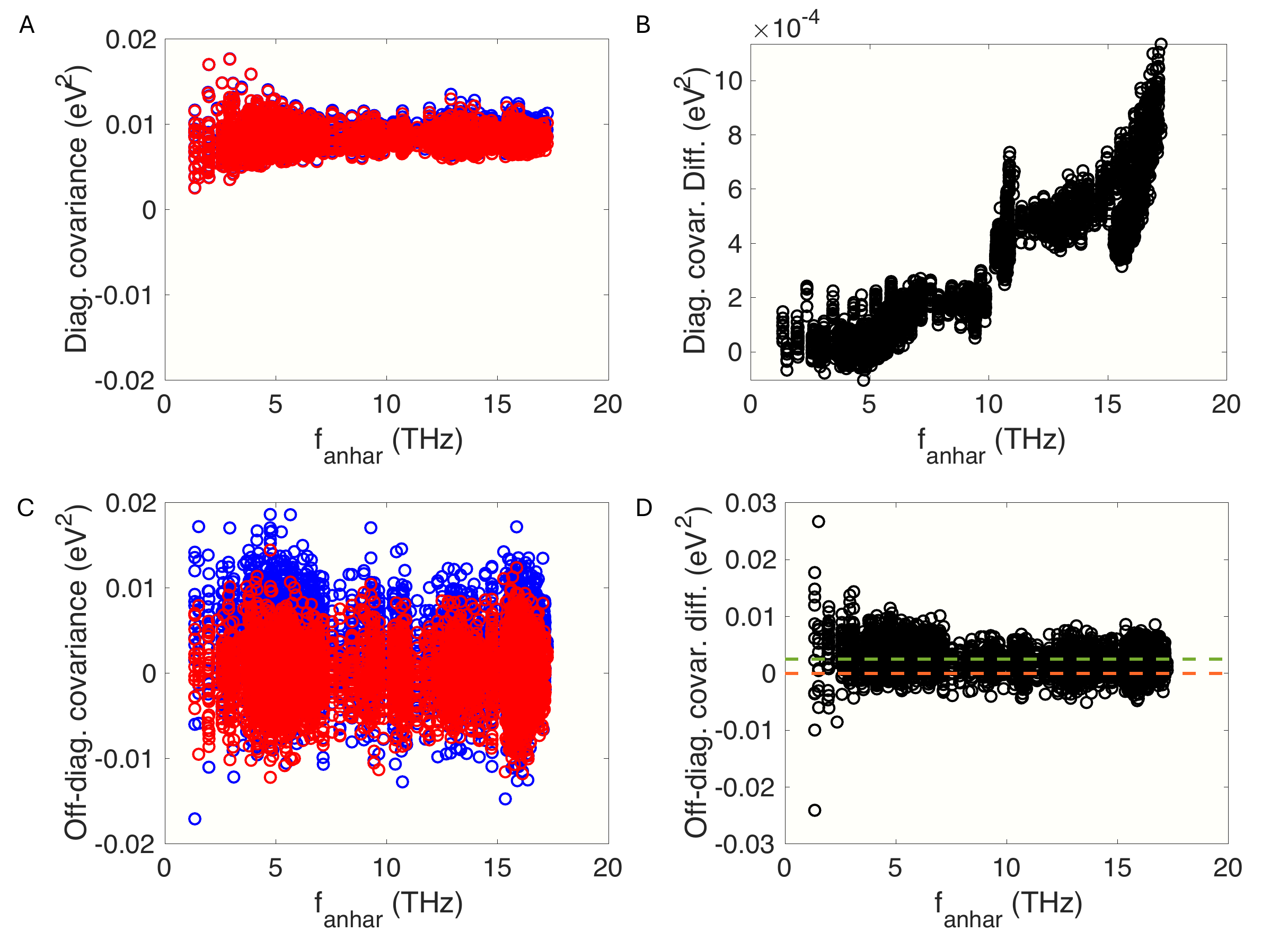}
	\caption{(A,C) Spectral self-variance and co-variance for mode potential (blue circles) and kinetic (red circles) energies, and (B, D) their differences (black circles) for 1500 K. For both (D), orange dashed lines represent zero and green dashed line is the actual mean.}
	\label{fig:S3}
\end{figure}

\begin{figure}[h!]
	\centering
	\includegraphics[width=1 \linewidth]{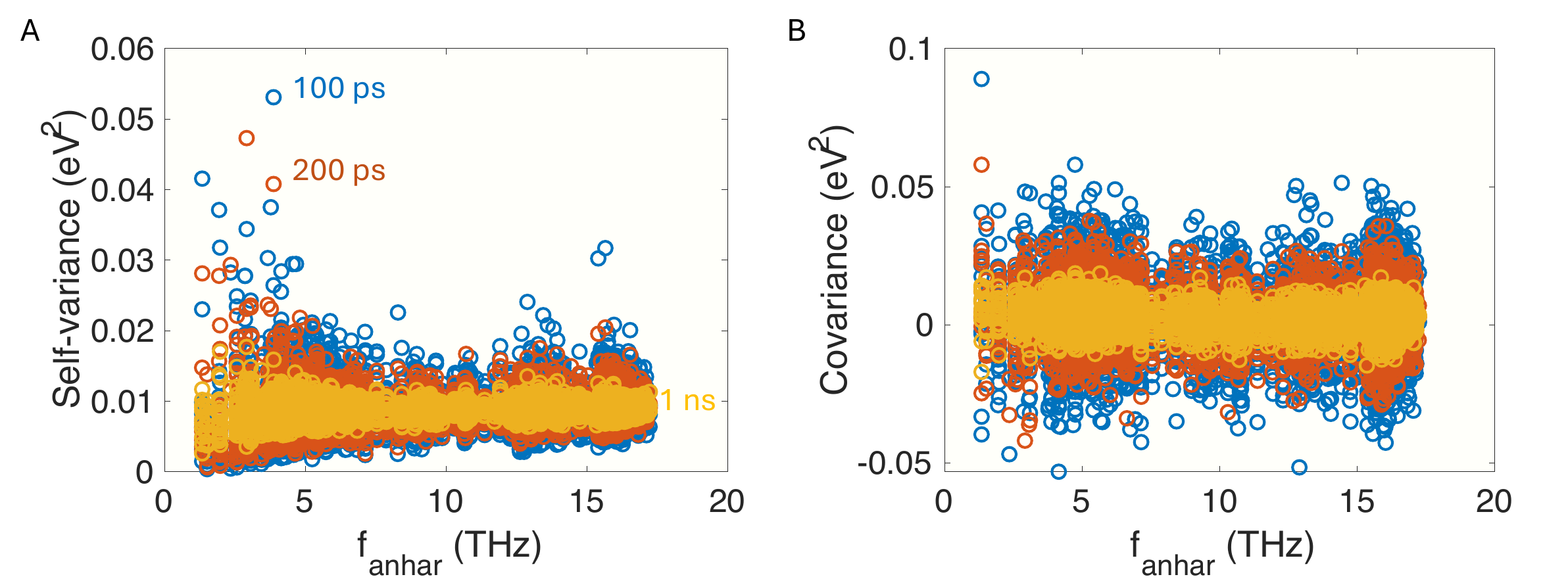}
	\caption{Time duration dependence (100 ps, 200 ps, and 1 ns) in (A) self-variance and (B) covariance of the mode potential energies at 1500 K. The spread of self-variance and co-variance becomes smaller with increase of time duration while the means stay nearly constant. Means of self-variance and co-variance are [0.0086, 0.0087, 0.0087] eV\textsuperscript{2} and [0.0029, 0.0028, 0.0028] eV\textsuperscript{2}, respectively. Time duration of 1 ns of data was used for the results discussed in the main text.}
	\label{fig:S5}
\end{figure}

\clearpage



\bibliography{aps.bib}

\end{document}